# Design and Analysis of Surface Plasmon Enhanced Electric Field in Al, Ag and Au Nanoparticle Dimers for UV-Visible Plasmonics

Soniya Juneja, and Manmohan Singh Shishodia*

Department of Applied Physics, School of Vocational Studies and Applied Sciences,
Gautam Buddha University, Greater Noida 201312, India

*Corresponding Author: **manmohan@gbu.ac.in**

**Abstract**

It is now well established that, due to excitation of localized surface plasmon resonances in plasmonic nanosystems, the electric field from an external irradiation source is greatly amplified in their vicinity. Surface plasmon mediated electric field amplification near plasmonic nanosystems is of vital importance for; light absorption enhancement of solar cells, molecular detection using surface enhanced Raman spectroscopy, proximity measurement of biological macromolecules such as proteins, lipids and nucleic acids using plasmon enhanced FRET, cancer therapy using photothermal effects, just to name only a few applications. Present work formulates Bergman-Milton multipole spectral expansion approach for investigating surface plasmon mediated electric field amplification in the dimer arrangement of Al, Ag and Au nanoparticles. In comparison to an isolated spherical shaped nanoparticle, their aggregate provide greater field enhancement, wider spectral tuning, and multiple resonant features. In view of the superior plasmonic characteristics of dimer configuration, a systematic investigation of electric field enhancement is done and their suitability for UV-Visible applications is assessed. The role of particle size, embedding medium and interparticle separation in field amplification and wavelength tuning is quantified. Moreover, plasmonic nanoparticle dimer designs for UV-Visible spectral window are presented.

I. **Introduction**

Surface plasmon (SP) mediated amplification and nanoscale confinement of electric field near plasmonic nanostructures plays a crucial role in enhancing light absorption [1-2], measuring proximity of biological macromolecules such as proteins, lipids and nucleic acids using SP enhanced FRET [3-4], altering electrical and optical properties of molecular nanojunctions [5], molecular detection using surface enhanced Raman spectroscopy (SERS) [6-7], photoluminescence [8-10], second harmonic generation [11], and fluorescence emission, to name only a few. It is well known that isolated (monomer) as well as coupled (*e.g.*, dimer) plasmonic nanoparticle (NP) system amplify external irradiation in their vicinity, largely due to excitation of localized surface plasmons (LSPs) and their coupling with irradiated field [12-24]. Generally, strongly coupled plasmonic dimer structures induce relatively stronger field amplification within the gap region of coupled nanoparticles. Interestingly, a molecule or a molecular chain placed in the dimer gap region allows a more efficient fluorescence quantum yield [22]. In fact, dimer system is superior in both, the field amplification as well as quantum yield, hence useful for enhancing fluorescence and photoluminescence. In contrast to the isolated nanoparticle, coupled nanoparticle structures in general provide greater flexibility in spectral tuning to obtain specifically engineered optical response at desired wavelength/frequency by controlling particle size and shape, particle separation, and plasmonic material. Localized surface plasmons (LSPs) of coupled nanoparticles further hybridize to produce bonding and antibonding plasmons, thus inducing a more intense electric field in the dimer gap relative to monomer structures. In the recent past, plasmonic properties of isolated and coupled nanoparticles have been studied extensively using *e.g.*, discrete dipole approximation (DDA) [13, 14], hybridization theory [13, 15], generalized multiparticle Mie (GMM) method [23], and FDTD [12] . For example, Hao and Schatz [14] studied EM fields and extinction efficiency around isolated and coupled nanoparticles of silver (Ag) using DDA for determining conditions to achieve field amplification sufficient for SERS based single molecule detection. After extensive theoretical and experimental investigations, the issue of the origin of enhanced field in SERS is settled and the major contribution to SERS enhancement is attributed to surface plasmon mediated local field amplification and subsequent enhancement of Raman excitation and emission near plasmonic nanoparticles. Optical properties of a pair of interacting identical gold nanoparticles were measured by **W. Rechberger** and co-workers [11], who reported a remarkable spectral red-shift for longitudinal polarization. **J. Aizpurua** and co-workers [15] used boundary element method (BEM) for studying optical response of coupled nanorods with varying composition, geometry, and environment with their prospective application in field enhanced spectroscopy. **K. H. Su** and co-workers **[13]** studied coupling between the dimer of elliptical gold NPs on quartz substrate through simulation and experiments. Li, Stockman and Bergman **[16]** proposed a self similar chain of Ag nanoparticles as plasmonic lens useful for optical detection, Raman characterization, and nanoparticle manipulation. It is now clear that, understanding the role of system geometry on plasmonic properties is indispensable for rational and optimal designs of plasmonic systems for specific application including but not limited to SERS, and biosensing. For external irradiated electric field $\mathbf{E}_0$, and resulting local electric field $\mathbf{E}$, the Raman enhancement factor is defined as, $\left(|\mathbf{E}|/|\mathbf{E}_0|\right)^4$. Also, the figure of merit of a bimolecular sensor is defined as, (FoM=Resonant Wavelength × Sensitivity/FWHM). Hence, it is desired to maximize the electric field enhancement for improved Raman signal, and to minimize (narrowed) the spectral width to improve the sensitivity of a bimolecular sensor. Due to recent surge of interests in photocatalysis and the detection of organic and biological systems showing strong UV response, aluminum based plasmonic system is gaining attention for their suitability in UV/deep-UV plasmonics [22, 23]. The renewed interest in Al plasmonics is primarily originated from the fact that, Al's bulk plasma frequency which lies in the UV region is higher than that of Au and Ag. This is useful for applications in, the detection of organic and biological systems exhibiting strong UV absorption, SERS substrate, plasmonic UV antenna, and photocatalysis.

Present study focus on formulating Bergman-Milton multipole spectral expansion method (**BMMSEM**) for investigating surface plasmon mediated electric field amplification in the dimer arrangement of Al, Ag and Au nanoparticles. **BMMSEM** was developed by D. J. Bergman for an accurate approximation of the effective dielectric constant of a two grain composite material system [24].

G. W. Milton further extended this method by giving generalization of the bounds [25]. Bergman spectral method has been recently extended to treat plexcitonic systems where energy transfer interactions between acceptor and donor molecules near isolated spherical nanoparticle and core-shell plasmonic nanoparticle [3-5]. Several easy to use closed form analytical expressions were developed, which were subsequently used for investigating surface plasmon enhanced strengthening of FRET process. **BMMSEM** separates geometrical and dielectric characteristics. Moreover, **BMMSEM** has flexibility and simplicity in extending the approach to treat an arbitrary arrangement of nanoparticles by solving corresponding general eigenvalue problem. The objectives of the present article are to: **(I).** formulate **BMMSEM** for calculating electric field near a dimer of spherical nanoparticles, **(II).** study the effect of particle size and interparticle separation on electric field enhancement, **(III).** compare the field enhancement characteristics of Ag, Al and Au as plasmonic materials and understand their behavior in UV-Vis regions, **(IV).** express basic dependence of electric field enhancement and the resonant peak on particle size, interparticle separation, and the embedding medium as fitting expressions handy to be used as basic design tools for designing nanoparticle dimer based plasmonic structures for UV-Visible spectral window useful for SERS, sensing etc. applications, **(V).** explain the physics behind important observations, **(VI).** propose Ag/Al/Au nanoparticle dimer designs at different wavelength.

## II. Theoretical Model

The dimer based plasmonic system of spherical nanoparticles (NPs) considered for present investigations is described in this section. The system consists of a homogenous pair of interacting metallic nanoparticles embedded in a medium of dielectric constant, $\varepsilon_d$ (see, Fig. 1). Left and right NPs are labeled as L and R, and their radii are labeled as $R_L$ and $R_R$, respectively. The optical response of constituent metallic nanoparticles is modeled using frequency dependent dielectric function $\varepsilon_m(\omega)$. Centre to centre and surface to surface separation between nanoparticles are denoted by b and S, respectively. The system is illuminated using radiation field polarized along dimer axis (here, z) and described as, $\mathbf{E} = E_0 \hat{\mathbf{z}}$, which corresponds to the external potential, $\Phi_{ext} = -E_0 z = -E_0 r \cos\theta$. The presence of plasmonic system alters local electric field in its vicinity, specifically in the gap region. The altered field can be useful for various applications including but not limited to SERS, molecular conduction and fluorescence. Here, we formulate Bergman-Milton multipole spectral expansion method (**BMMSEM**) to calculate electric field enhancement in the gap region between coupled spherical nanoparticles and investigate the properties of Al, Ag, and Au based dimer configuration. In the dimer system consisting of two spherical nanoparticles (SNPs) having separation b, the centre of left NP is considered at the origin of the coordinate system. The centre of right NP will always be shifted relative to the origin. To analyze such a plasmonic system with respect to a common origin, the eigenstates of right NP will have to be expressed relative to the adopted common origin, here $O_L$. This requires the transformation of eigenstates of right NP. In such a case, the function consisting of multipoles has to be expanded about the shifted origin. The optical response of the system is described through dielectric function $\varepsilon(\mathbf{r}, \omega)$ as [4],

$$\varepsilon(\mathbf{r},\omega) = \varepsilon_m(\omega)\,\Theta(\mathbf{r}) + \varepsilon_d[1-\Theta(\mathbf{r})] \qquad [1]$$

where, $\Theta(\mathbf{r}) = 1\,(0)$ inside (outside) the metal. In the limits of quasistatic approximation, where considered system size is much less than the wavelength ($\lambda$) of light, the electrostatic potential $\Phi(\mathbf{r})$ can be expressed as [4],

$$\nabla \bullet [\varepsilon(\mathbf{r},\omega)\nabla\Phi(\mathbf{r})] = 0 \qquad [2]$$

Using Eq. 1, Eq. 2 produces,

$$\nabla \bullet \{\varepsilon_m(\omega)\Theta(\mathbf{r}) + \varepsilon_d[1-\Theta(\mathbf{r})]\}\nabla\Phi(\mathbf{r}) = 0 \qquad [3]$$

Further mathematical manipulation leads to,

$$\nabla^2 \Phi(\mathbf{r}) = [1 - \{\varepsilon_m(\omega)/\varepsilon_d\}][\nabla \bullet \Theta(\mathbf{r})\nabla\Phi(\mathbf{r})] \qquad [4]$$

Following references [4, 24] define spectral parameters, $u(\omega)$ and $s(\omega)$ as,

$$s(\omega) = [1 - \{\varepsilon_m(\omega)/\varepsilon_d\}]^{-1}, \quad u(\omega) = 1/s(\omega) \qquad [5]$$

Eq. 4 in terms of spectral parameters becomes,

$$\nabla \bullet [\Theta(\mathbf{r}) \nabla \Phi(\mathbf{r})] = s(\omega) \nabla^2 \Phi(\mathbf{r}) \qquad [6]$$

The corresponding eigenvalue problem now becomes,

$$\nabla \bullet [\Theta(\mathbf{r}) \nabla \psi_\lambda(\mathbf{r})] = s_\lambda \nabla^2 \psi_\lambda(\mathbf{r}) \qquad [7]$$

In general, one will have to solve Eq. 6 (equivalently, Eq. 7) for the plasmonic system under consideration. The formal solution for $\Phi(\mathbf{r})$ may be written in terms of vacuum Green's function $G_0$ as [24],

$$\Phi(\mathbf{r}) = \Phi_{ext}(\mathbf{r}) - u(\omega) \iiint d^3 r' G_0(\mathbf{r}, \mathbf{r}') \nabla' \bullet [\Theta(\mathbf{r}') \nabla' \phi(\mathbf{r}')] \qquad [8]$$

$$= \Phi_{ext}(\mathbf{r}) + u(\omega) \Gamma \Phi(\mathbf{r}) \qquad [9]$$

where, operator $\Gamma$ is defined as the following integral,

$$\Gamma \Phi(\mathbf{r}) = \iiint d^3 r' \Theta(\mathbf{r}') [\nabla' G_0(\mathbf{r}, \mathbf{r}')] \nabla' \phi(\mathbf{r}') \qquad [10]$$

Here, the scalar product of two eigenfunctions is defined as [29],

$$\langle \psi_\lambda(\mathbf{r}) | \psi_{\lambda'}(\mathbf{r}) \rangle = \iiint_V \Theta(\mathbf{r}) \nabla \psi_\lambda^*(\mathbf{r}) \nabla \psi_{\lambda'}(\mathbf{r}) d^3 r = \delta_{\lambda, \lambda'} \qquad [11]$$

Here, $\psi_\lambda^*(\mathbf{r})$ denotes the complex conjugated and $\delta$, the Kronecker delta symbol. The scalar product is valid inside the metallic region only. When no external potential is applied, the eigenvalue equation takes the following form,

$$s_\lambda \psi_\lambda(\mathbf{r}) = \Gamma \psi_\lambda(\mathbf{r}) \qquad [12]$$

The detailed procedure for determining eigenvalues and eigenstates of operator $\Gamma$ for isolated spherical and core-shell nanoparticle is described in [3, 4]. On determining the eigenvalues and eigenstates of operator $\Gamma$, the solution can now be generalized to the external electric potential, say $\Phi_{ext}(\mathbf{r})$. Using the completeness condition,

$$I = \sum_{\lambda} |\psi_\lambda(\mathbf{r})\rangle \langle \psi_\lambda(\mathbf{r})| \qquad [13]$$

The potential $\Phi(\mathbf{r})$ at point $\mathbf{r}$ due to the dimer system can be written as [24],

$$\Phi(\mathbf{r}) = \Phi_{ext}(\mathbf{r}) + \sum_{\lambda} \sum_{\substack{u,l \\ u=L,R}} \sum_{\substack{v,l' \\ v=L,R}} \frac{s_{vl'}}{s(\omega) - s_\lambda} (B_{ul}^\lambda)(B_{vl'}^\lambda) \langle \Theta_u^+ \psi_{a,l} | \Phi_{ext} \rangle \psi_{v,l'}(\mathbf{r}) \qquad [14]$$

The electric potential in the gap region is calculated using Eq. 14 and subsequently the electric field in the middle of the dimer gap by taking gradient of the electric potential. Now, for calculating the overall potential in the gap region, firstly one will have to calculate the external potential due to the irradiated homogeneous electric field in which the whole dimer system is kept. For this, one will have to; construct dimer matrix, calculate eigenvalues and eigenvectors of dimer matrix, obtain translated eigenmodes, and solve overlap integrals for translated and origin centered nanoparticle.

**II.A.    Dimer Matrix Element**

As discussed above, the system consists of a pair of spherical nanoparticles, namely 'L' and 'R', their radii denoted as $R_L$ and $R_R$, where L is centered at the origin and R is centered at, $\mathbf{b} = (b, \theta_b, \phi_b)$. The eigenstates and eigenvalues of dimer will be found using the union of eigenstates of each individual nanoparticle, which is defined as [24],

$$\Gamma = \begin{bmatrix} \Gamma_{L,l\,m;\,L,l'\,m'} & \Gamma_{L,l\,m;\,R,l'\,m'} \\ \Gamma_{R,l\,m;\,L,l'\,m'} & \Gamma_{R,l\,m;\,R,l'\,m'} \end{bmatrix} \qquad [15]$$

where, matrix element $\Gamma_{L,lm;\,R,l'm'}$ is defined as,

$$\Gamma_{lm;\,l'm'} = \frac{l'}{2l'+1} \int_V d^3r\,\theta_L \left[ \nabla \psi_{lm}^* \cdot \nabla \psi_{l'm'}(\mathbf{r}-\mathbf{b}) \right] \qquad [16]$$

Here, $\psi_{lm}$ ($\psi_{l'm'}$) corresponds to the inner (outer) eigenstates of L (R) nanoparticle. Using Green's identity, the integral can be simplified as,

$$\int_V d^3r [\nabla \Psi_{lm}^* \cdot \nabla \Psi_{l'm'}(\mathbf{r}-\mathbf{b})] = \oint_S R_L^2\, d\Omega\, \psi_{l'm'}(\mathbf{r}-\mathbf{b}) \frac{\partial \Psi_{lm}^*(\mathbf{r})}{\partial r} - \int_V d^3r\, \theta_L [\Psi_{l'm'}(\mathbf{r}-\mathbf{b}) \nabla^2 \Psi_{lm}^*(\mathbf{r})] \qquad [17]$$

Using Eq. 17, and the property of generalized Laplace equation, $\nabla^2 \Psi^*_{L;lm} = 0$, Eq. 16 produces,

$$\Gamma_{lm,l'm'} = \frac{l'}{2l'+1} \oint_{r=R_L} R_L^2 d\Omega \frac{\partial \Psi^*_{lm}(\mathbf{r})}{\partial r} \Psi_{l'm'}(\mathbf{r}-\mathbf{b}) \qquad [18]$$

As derived elsewhere, inner eigenmodes of origin centered spherical nanoparticle are,

$$\Psi_{lm}(\mathbf{r}) = \frac{1}{\sqrt{lR_L^{2l+1}}} r^l Y_{lm}(\theta,\phi) \qquad [19]$$

Therefore,

$$\frac{\partial \Psi^*_{lm}(\mathbf{r})}{\partial r} = \frac{1}{\sqrt{lR_L^{2l+1}}} r^{l-1} Y^*_{lm}(\theta,\phi) \qquad [20]$$

Similarly, the outer eigenmodes of right nanoparticle are,

$$\Psi_{l'm'}(\mathbf{r}-\mathbf{b}) = \sqrt{\frac{R_R^{2l'+1}}{l'}} \frac{1}{|\mathbf{r}-\mathbf{b}|^{l'+1}} Y_{l'm'}(\Omega_{r-b}) \qquad [21]$$

The term, $\frac{1}{|\mathbf{r}-\mathbf{b}|^{l'+1}} Y_{l'm'}(\Omega_{r-b})$ using [24] can be written as,

$$\frac{1}{|\mathbf{r}-\mathbf{b}|^{(l'+1)}} Y_{l'm'}(\Omega_{r-b}) = \sum_{\mu\lambda} (-1)^{m'+l'} \sqrt{4\pi(2l'+1)[2(\lambda+l')+1](2\lambda+1)} \begin{pmatrix} l' & l'+\lambda & \lambda \\ 0 & 0 & 0 \end{pmatrix}$$

$$\times \begin{pmatrix} l' & l'+\lambda & \lambda \\ -m' & m'-\mu & \mu \end{pmatrix} \frac{[2(\lambda+l')-1]!!}{(2\lambda+1)!!(2l'-1)!!} \frac{r^\lambda}{b^{\lambda+l'+1}} Y_{\lambda,\mu}(\theta,\phi) Y_{\lambda+l',m'-\mu}(\theta_b,\phi_b)$$

Therefore,

$$\Psi_{l'm'}(\mathbf{r}-\mathbf{b}) = \sqrt{\frac{R_R^{2l'+1}}{l'}} \sum_{\mu\lambda} (-1)^{m'+l'} \sqrt{4\pi(2l'+1)[2(\lambda+l')+1](2\lambda+1)} \begin{pmatrix} l' & l'+\lambda & \lambda \\ 0 & 0 & 0 \end{pmatrix}$$

$$\times \begin{pmatrix} l' & l'+\lambda & \lambda \\ -m' & m'-\mu & \mu \end{pmatrix} \frac{[2(\lambda+l')-1]!!}{(2\lambda+1)!!(2l'-1)!!} \frac{r^\lambda}{b^{\lambda+l'+1}} Y_{\lambda,\mu}(\theta,\phi) Y_{\lambda+l',m'-\mu}(\theta_b,\phi_b) \qquad [22]$$

Using Eqs. 20-22, the matrix element in Eq. 18 becomes,

$$\Gamma_{lm;l'm'} = \frac{l'}{2l'+1} \oint_{S(r=R_L)} R_L^2 d\Omega \frac{l}{\sqrt{lR_L^{2l+1}}} r^{l-1} Y_{lm}^*(\theta,\phi) \sqrt{\frac{R_R^{2l'+1}}{l'}} \sum_{\mu\lambda} (-1)^{m'+l'} \sqrt{4\pi(2l'+1)[2(\lambda+l')+1](2\lambda+1)}$$

$$\times \begin{pmatrix} l' & l'+\lambda & \lambda \\ 0 & 0 & 0 \end{pmatrix} \begin{pmatrix} l' & l'+\lambda & \lambda \\ -m' & m'-\mu & \mu \end{pmatrix} \frac{[2(\lambda+l')-1]!!}{(2\lambda+1)!!(2l'-1)!!} \frac{r^\lambda}{b^{\lambda+l'+1}} Y_{\lambda,\mu}(\theta,\phi) Y_{\lambda+l',m'-\mu}(\theta_b,\phi_b)$$

$$\Gamma_{lm;l'm'} = (-1)^{m-l'} \left(\frac{R_L}{b}\right)^{l+(1/2)} \left(\frac{R_R}{b}\right)^{l'+(1/2)} \sqrt{\frac{ll'}{(2l+1)(2l'+1)}} \frac{(l+l'+m-m')!}{\sqrt{(l+m)!(l-m)!(l'-m')!(l'+m')!}} P_{l+l'}^{m'-m} e^{i(m'-m)\phi_b}$$

[23]

## II.B. Overlap Integrals

In this section, we present the procedure to evaluate integrals like $\langle \Theta \psi_{a,l} | \Phi_{ext} \rangle$ appearing in Eq. 14.

As described above, the overlap integral, $\langle \theta_L \Psi_{L;lm} | \Phi_{ext} \rangle$ is defined as,

$$\langle \theta_L \Psi_{L;lm} | \Phi_{ext} \rangle = \int_V d^3r \, \theta_L \left[ \nabla \Psi_{L;lm}^* \cdot \nabla \Phi_{ext} \right] \quad [24]$$

Here, we are considering the case of left nanoparticle, 'L', where $\Psi_{L,lm}$ denotes the eigenstates for particle L and $\theta_L$ signify the validity of scalar product only inside nanoparticle. Now, let's make use of the Green's first identity valid for two arbitrary scalar fields $\phi$ and $\psi$ [28]:

$$\int_V \left( \phi \nabla^2 \psi + \nabla \phi \cdot \nabla \psi \right) d^3x = \oint_S \phi \frac{\partial \psi}{\partial n} dA$$

where, $(\partial \psi / \partial n) dA = \nabla \psi \cdot \hat{n}$, and $\hat{n}$ corresponds to the normal, outward to surface S. Hence,

$$\int_V d^3r \, \theta_L \left[ \nabla \Psi_{L;lm}^* \cdot \nabla \Phi_{ext} \right] = \oint_S dA \left[ \Phi_{ext} \nabla \Psi_{L;lm}^* \cdot \hat{n} \right] - \int_V d^3r \, \theta_L \left[ \Phi_{ext} \nabla^2 \Psi_{L;lm}^* \right]$$

The closed surface integral is performed over the surface of the nanoparticle. The volume integral in RHS of above equation vanishes by virtue of the property of Laplace equation. Therefore,

$$\langle \theta_L \Psi_{L;lm} | \Phi_{ext} \rangle = \oint_S dA \left[ \Phi_{ext} \nabla \Psi_{L;lm}^* \cdot \hat{n} \right] \quad [25]$$

For isolated nanoparticle, L of radius $R_L$, the eigenfunctions outside the particle at position, $\mathbf{r} = (r, \theta, \phi)$ were derived in [4] as, $\Psi_{L;lm} = Y_{lm}(\theta, \phi) r^l / \sqrt{l R_L^{2l+1}}$

Equation 25 can now be written as,

$$\langle \theta_L \Psi_{L;lm} | \Phi_{ext} \rangle = -E_0 R_L^{3/2} \sqrt{l} \oint_S \cos\theta \, Y_{lm}^*(\theta, \phi) \sin\theta \, d\theta \, d\phi \qquad [26]$$

Now using properties of spherical harmonics and orthogonality relation [28], we can write,

$$\langle \theta_L \Psi_{L;lm} | \Phi_{ext} \rangle = -\sqrt{\frac{4\pi}{3}} E_0 R_L^{3/2} \qquad [27]$$

Other overlap integrals are calculated following the similar procedure. In our implementation, matrix elements, overlap integrals with external field, eigenmodes of individual particles inside and outside are calculated analytically using closed form expressions. We developed our own code in **MATLAB 7.10.0 (R2010a)**. Matrix manipulation such as determination of eigenvalues and eigenfunctions of matrix-$\Gamma$ is done numerically using built-in MATLAB commands. The electric field at a point is calculated by taking the gradient of the potential *i.e.*, $\mathbf{E} = -\nabla\Phi$.

### III. Results and Discussion

This section provides the results of our investigations based on theoretical approach described in Sec. **II**. Present study considers plasmonic nanoparticles (PNPs) made of aluminum ($_{13}$Al), silver ($_{47}$Ag), and gold ($_{79}$Au) in isolated as well as dimer configuration. Radius of each constituent nanoparticle is denoted by R, and the interparticle separation (surface to surface) is denoted by S. Unless mentioned otherwise, both PNPs in the dimer system will be considered spherical in shape and identical in size (R) and composition ($\varepsilon_m$). The optical response of PNPs is described through photon energy ($\hbar\omega$) dependent optical constants of Johnson and Christy for Au and Ag [26], and A. D. Rakic for Al [27]. We expressed the optical constants from these references into fitting expressions, whose validity is shown by comparing the fitted (-, solid line), and the actual data (o, symbol) for Al, Ag, and Au in Fig. 2. The quality factor defined as, $Q = [-\varepsilon_m'(\omega) / \varepsilon_m''(\omega)]$ is shown as an inset in Fig. 2(b) for Al, Au and Ag. Evidently, the calculated variation of Q shows excellent match with the literature [29]. High Q value corresponds to strong confinement of plasmons in the cavity and large number of bounces inside the cavity before escaping. Figure 2 show that Ag possesses highest Q-factor and Al the lowest Q-factor in Vis-NIR region. This indicates lowest (highest) rate of energy loss relative to the stored energy for Ag (Al) in the Vis-NIR spectral region. However, one can notice that, the Q-factor for Al is greater in UV and deep UV (DUV) spectral region. Thus Ag shows best plasmonic properties in Vis-NIR region and Al in UV-DUV spectral region. Quantitatively speaking, the maximum Q-factor for Ag, Al and Au corresponds to $\hbar\omega$ =1.52 eV ($\lambda$~815 nm), $\hbar\omega$ =7.58 eV ($\lambda$~163 nm), and $\hbar\omega$ =1.14 eV ($\lambda$~1087 nm), respectively. However, Q-factor for Al exceeds that of Au and Ag for $\hbar\omega$>2.26 eV ($\lambda$<550 nm) and 3.56 eV ($\lambda$<350 nm), respectively. The field amplification factor $A(\omega) = \left|\frac{\mathbf{E(r)}}{\mathbf{E}_0}\right|^2$ as a function of $\hbar\omega$ for isolated and coupled spherical nanoparticles of Ag, Al and Au is presented in Fig. 3. The variation of $A(\omega)$ for isolated SNPs of Ag, Al and Au at observation point r=1.0-nm away from the particle surface with varying nanoparticle radius (R) =10, 20, 30 and 35 nm, is presented in Fig. 3(a). The field amplification factor is calculated using, $A(\omega) = \left|\frac{\mathbf{E(r)}}{\mathbf{E}_0}\right|^2 = \left|1 + \frac{2R^3}{r^3}\frac{\varepsilon_m(\omega) - \varepsilon_d}{\varepsilon_m(\omega) + 2\varepsilon_d}\right|^2$. Figure 3(b-d) shows the variation of $A(\omega)$ for dimer configuration with, R=10, 20, and 35 nm, using **BMMSEM** is discussed in Sec. II. System is assumed to be embedded in a medium with $\varepsilon_d$=2.5 and the interparticle separation, S=1.0-nm is considered. Isolated nanoparticle shows field amplification peak at 6.45 eV/192.24 nm (Al), 3.02 eV/410.59 nm (Ag), and 2.24 eV/553.57 nm (Au) where corresponding amplification factor $A(\omega)$ at peak values are, $1.07\times10^3$, $3.25\times10^3$, and $5.47\times10^1$, respectively. This may be noticed that the spectrum of isolated nanoparticle exhibits only a single resonant peak, whose position is quite insensitive to the particle size within the quasistatic size limit. Moreover, field enhancement factor produced by isolated nanoparticles is normally insufficient for SERS based single molecule detection [30]. The main features of Fig. 3 may be summarized as: **(I).** isolated particle shows $A(\omega)$ ~$10^3$ while dimer spectra shows $A(\omega)$ ~$10^6$, **(II).** in contrast to a single peak for isolated NP, dimer configuration shows multiple resonant peaks, **(III).** the number of plasmon peaks increases with NP size, for example Al shows three peaks at R=10-nm but four peaks at R=35 nm, **(IV).** Au shows less number of peaks and smaller A(ω) relative to Ag, **(V).** A(ω) increases significantly with R, **(VI).** peak separation for Al is more than Ag and Au. Similar calculations were done for different sizes and for the system embedded in the mediums of different permittivities (not shown for brevity). The results for the first plasmon peak are summarized in Fig. 4 (a-d). Also shown is the fitted data for all cases. Resonant peak positions ($\lambda_R$) for all three metals show redshift on increasing particle size (R). The analysis shows that the redshift can be fitted as a linear function of R as,

$$\lambda_R (nm) = \lambda_0 (nm) + a\ R\ (nm) \qquad [28].$$

where, $\lambda_0$, and $a$ are the fitting parameters. Here $\lambda_R$ and R are expressed in nanometer units. The details of the fitted parameters for $\varepsilon_d$=1, 2, 3 and 4 are provided in Table-1. Evidently, the rate of redshift is maximum for Ag. However, for low dielectric constant medium, rate of red-shift is more for Al, while Au supersedes this for embedding medium with high dielectric constant. Clearly, the plasmon peaks can be tuned by controlling the particle size in the dimer configuration. The amplification factor, **A(ω)** as a function of photon energy (ℏω) for coupled SNPs of Al, Ag and Au is shown in Fig. 5 for different embedding mediums characterized by, $\varepsilon_d$=1, 2, 3, and 4 for fixed particle separation, S=1.0-nm and particle size, R=20-nm. The most important features may be summarized as: **(I).** A(ω) increases with increasing dielectric constant $\varepsilon_d$ of embedding medium, **(II).** more peaks appear in the spectra with increasing dielectric constant $\varepsilon_d$ of embedding medium, **(III).** spectra as well as resonant peak positions get shifted towards low photon energy (redshift). Calculated dependence of resonant positions (first peak) $\lambda_R$ on the permittivity of embedding medium $\varepsilon_d$ for varying radii, R=10, 20, 30 and 35-nm is shown in Fig. 6. For all calculations, S=1.0 nm is considered. It can be seen that, the resonant position $\lambda_R$ gets redshifted with increasing $\varepsilon_d$. The behavior suggests that, $\lambda_R$ can be fine tuned through the selection of embedding medium ($\varepsilon_d$) and the variation can be fitted into the following mathematical expression,

$$\lambda_R (nm) = \lambda_0 (nm) + a\ \varepsilon_d \qquad [29].$$

where, $\lambda_0$ and $a$ are fitting coefficients, and $\lambda_R$ is expressed in nanometer units. The details of fitting parameters are provided in **Table-2.** The effect of interparticle separation S on $\lambda_R$ is shown in Fig. 7. Evidently, resonant position gets blueshifted on increasing interparticle separation S. The blueshift with interparticle separation can be fitted as,

$$\lambda_R (nm) = \lambda_0 (nm) + a_0\, e^{-S/S_0} \qquad [30].$$

where, $\lambda_0$, $a_0$ and $S_0$ are fitting coefficients and all are expressed in nanometer units. The details of the fitting parameters are provided in Table-3 (a). Also, the effect of interparticle separation S on field amplification factor, A ($\omega$) is shown in Fig. 8. Amplification factor, A ($\omega$) increases for all three metals on reducing interparticle separation. This is due to the fact that, on increasing interparticle separation S, the interaction between constituting NPs will be weakened and fewer modes will get coupled. The variation of field amplification factor, A ($\omega$) with S (nm) can be fitted in the following expression,

$$ln\ (\mathbf{A}) = a_0 - a_1\ ln\ (S) \qquad [31].$$

where, $a_0$ and $a_1$ are fitting coefficients and S is expressed in nm units. The details of the fitting parameters are provided in **Table-3(b).** The mathematical fitting expressions provide simple tools for understanding the behavior of the dimer based plasmonic systems and designing plasmonic system resonating at desired wavelength of interest in UV-Vis spectral region. In order to show the significance of fitting expressions developed in this work, we use them to design dimer based plasmonic system with maximum field enhancement at six important wavelengths of practical importance [30], $\lambda$=266, 315, 488, 514, 633, and 660 nm.

## IV.    Conclusion

In conclusion, we have formulated and used Bergman-Milton multipole spectral expansion method to investigate the electric field enhancement in the gap region of Al, Ag and Au nanoparticle dimers. The role of particle size, embedding medium, interparticle separation and the metal type are studied in details. The resonant wavelength shows redshift with increasing nanoparticle radius as well as the dielectric constant of the embedding medium. The redshift with increasing particle size and the dielectric constant of embedding medium has been expressed as a linear fitting expression. The blueshift of resonant wavelength on increasing interparticle separation is fitted as single phase exponential decay. Moreover, the field amplification factor and interparticle separation are fitted into log-log expression for Al, Ag and Au. Most importantly, these fitting relations are used to design plasmonic nanoparticle dimer with resonant optical response at crucial wavelengths in UV-Vis spectral region.

**Acknowledgments**

M. S. Shishodia is pleased to acknowledge **Dr. M. Oren** and **Professor A. Nitzan** (School of Chemistry, Tel Aviv University, Israel), and **Professor B. D. Fainberg** (H.I.T, Israel) for their help, support and illuminating discussions.

**Table-1:** The details of the fitting parameters for R (nm) dependence of $\lambda_R$ (nm) for different dielectric constant ($\varepsilon_d$) of embedding medium and interparticle separation, S=1.0 nm.

| Metal | $\varepsilon_d$=1.0 | | $\varepsilon_d$=2.0 | | $\varepsilon_d$=3.0 | | $\varepsilon_d$=4.0 | |
|---|---|---|---|---|---|---|---|---|
| | $\lambda_0$ | a | $\lambda_0$ | a | $\lambda_0$ | a | $\lambda_0$ | a |
| Al | 165.18 | 2.04 | 224.59 | 2.99 | 273.15 | 3.43 | 312.65 | 3.75 |
| Ag | 375.96 | 2.60 | 444.69 | 4.42 | 506.83 | 6.06 | 580.48 | 6.47 |
| Au | 530.45 | 1.55 | 564.27 | 4.00 | 620.21 | 5.40 | 675.24 | 6.42 |

**Table 2:** The details of the fitting parameters for $\varepsilon_d$ dependence of $\lambda_R$ for different particle radii (R) and interparticle separation, S=1.0 nm.

| Metal | R=10 nm | | R=20 nm | | R=30 nm | | R=35 nm | |
|---|---|---|---|---|---|---|---|---|
| | $\lambda_0$ | *a* | $\lambda_0$ | *a* | $\lambda_0$ | *a* | $\lambda_0$ | *a* |
| Al | 137.97 | 53.20 | 159.08 | 60.66 | 175.57 | 65.057 | 182.81 | 66.61 |
| Ag | 325.29 | 77.44 | 341.79 | 95.97 | 358.50 | 107.45 | 365.47 | 111.93 |
| Au | 478.36 | 62.34 | 479.71 | 83.501 | 485.45 | 97.17 | 489.47 | 102.84 |

**Table-3:** The details of the fitting parameters for interparticle separation, S dependence of $\lambda_R$ and A for nanoparticle radius, R=20 nm and the dielectric constant of embedding medium, $\varepsilon_d$=2.5.

| (a). R=20 nm, $\varepsilon_d$=2.5 $\lambda_R (nm) = \lambda_0 + a_0 e^{-S/S_0}$ | | | | (b). R=20 nm, $\varepsilon_d$=2.5 $ln(A) = a_0 - a_1 ln(S)$ | |
|---|---|---|---|---|---|
| **Metal** | $\lambda_0$ | $a_0$ | $S_0$ | $a_0$ | $a_1$ |
| **Al** | 216.17 | 154.38 | 2.25 | 10.61 | 1.59 |
| **Ag** | 442.88 | 222.47 | 2.07 | 13.78 | 2.01 |
| **Au** | 572.56 | 213.93 | 1.617 | 12.14 | 3.31 |

**Table-4:** Design parameters for six important wavelengths in UV-Visible region for nanoparticle size, R=20 nm and embedding medium dielectric constant, $\varepsilon_d$=2.5.

| λ (nm) | Interparticle Separation (nm) | Amplification Factor | Nanoparticle Material |
|---|---|---|---|
| 266 | 2.54 | $00.90 \times 10^4$ | Al |
| 315 | 1.00 | $04.03 \times 10^4$ | Al |
| 488 | 3.30 | $08.74 \times 10^4$ | Ag |
| 514 | 2.36 | $17.17 \times 10^4$ | Ag |
| 633 | 2.04 | $01.76 \times 10^4$ | Au |
| 660 | 1.45 | $05.51 \times 10^4$ | Au |

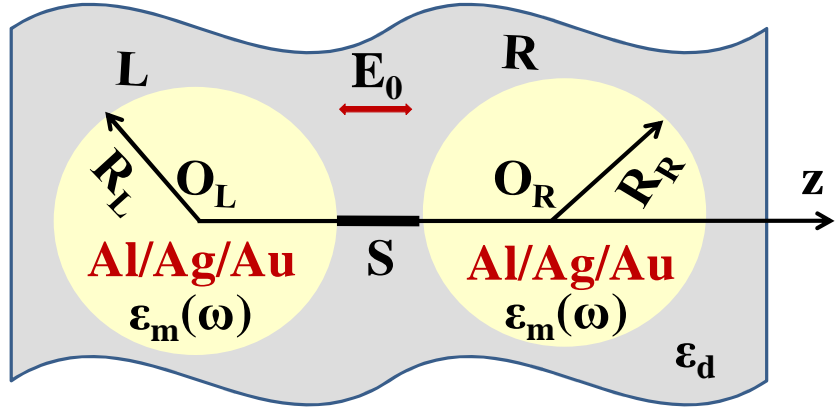

**Fig. 1:** The schematic of dimer system of coupled spherical nanoparticles (SNPs) under consideration. Here, left and right NPs are labeled as L and R, their centre as $O_L$ and $O_R$ and their radii as $R_L$ and $R_R$, respectively. Both NPs are considered embedded in a homogeneous dielectric medium ($\varepsilon_d$). The dimer system is exposed to homogenous external electric field along **z**-direction which excites localized surface plasmons (LSPs) of NPs. Interparticle separation (S) dependent hybridization of these LSPs leads to the formation of bonding (lower energy) and antibonding plasmons (higher energy) which subsequently produce large field enhancement in the dimer gap. The optical response of metallic nanoparticles is taken into account through $\hbar\omega$ dependent dielectric function. Unless mentioned otherwise, both NPs will be considered identical (size, shape and composition) for calculating field enhancement at midpoint of the gap. Surface to surface distance (hereafter referred to as interparticle separation) is denoted by S. The system is illuminated using external irradiation polarized along dimer axis.

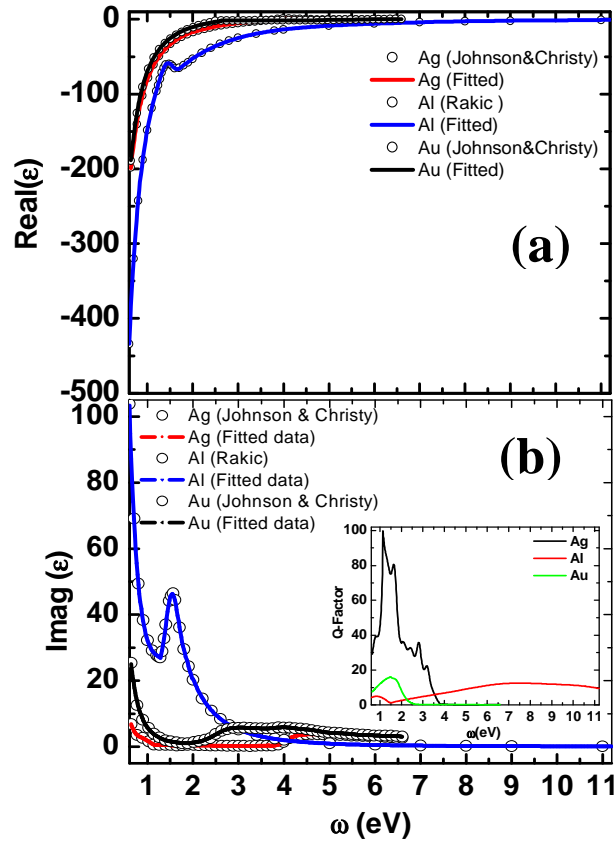

**Fig. 2:** The real (a) and the imaginary part (b) of dielectric function (o-symbol) for Al, Ag and Au metals based on Johnson and Christy [26] for Au & Ag and on A. D. Rakić (27) for Al. The fitted data is shown by solid lines (-) for all three plasmonic materials. The quality factor Q for Ag, Al and Au is plotted in the insert of Fig. 2(b). A high Q factor means that the plasmons are strongly confined and bounce around many times inside the cavity before escaping. Highest quality factor Q for Ag in visible region indicates lowest rate of energy loss relative to the stored energy for Ag and highest rate of energy loss for Al. However, Q factor for Al is highest in UV and deep UV region. Thus Ag shows best plasmonic properties in visible region and Al in the UV region.

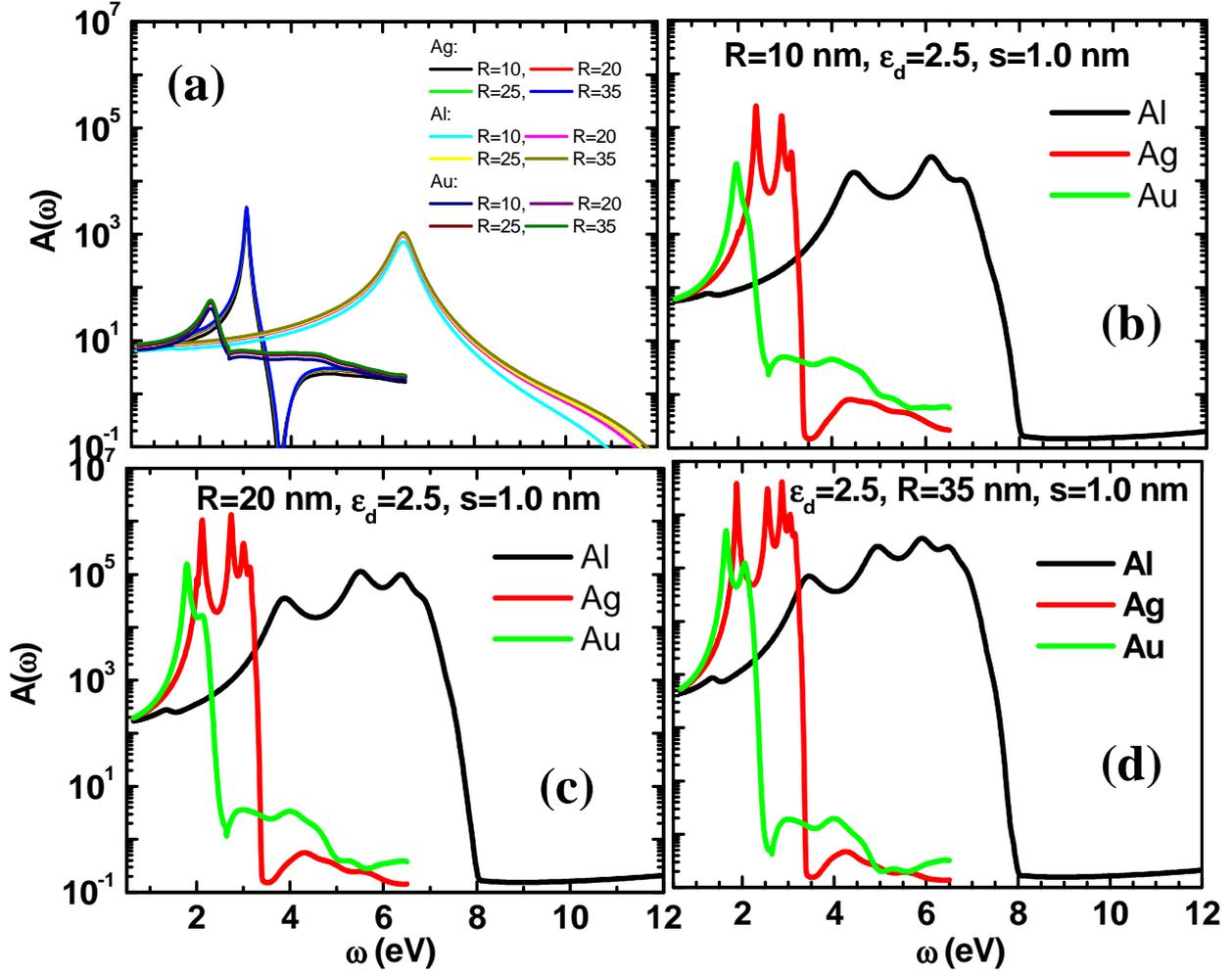

**Fig. 3:** Amplification factor A(ω) as a function of ℏω for isolated (a), and dimer configuration of SNPs of Al, Ag and Au and varying nanoparticle radii, R=10 **(b)**, 20 **(c)** and 35 **(d)**. Both, isolated and coupled nanoparticles are considered to be embedded in a medium of dielectric constant $\varepsilon_d$=2.5 and interparticle separation in dimer, S=1.0 nm. Isolated nanoparticle shows peaks at, ℏω=2.24 eV (λ=553.57 nm), ℏω=3.02 eV (λ=410.59 nm), and ℏω=6.45 eV (λ=192.24 nm) for Au, Ag and Al respectively. It is evident that the dimer structures can produce substantially higher field enhancement in the gap region compared to the isolated nanoparticle.

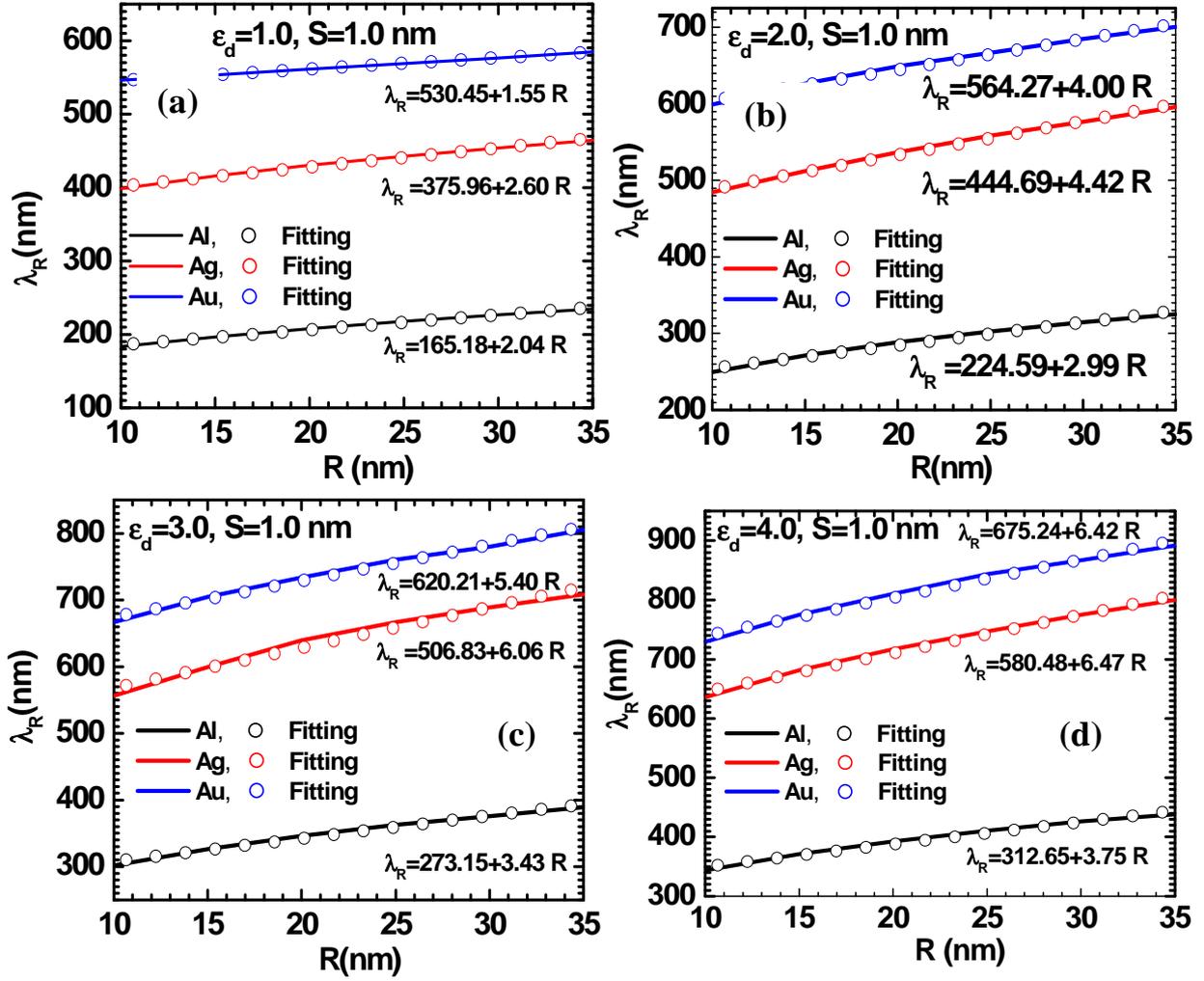

**Fig. 4:** The calculated variation of resonant wavelength $\lambda_R$ with nanoparticle radii R (both NPs are considered identical) for Al, Ag and Au plasmonic materials embedded in the medium with $\varepsilon_d$=1.0 (a), $\varepsilon_d$=2.0 (b), $\varepsilon_d$=3.0 (c) and $\varepsilon_d$=4.0 (d). The variation is fitted as a straight line. Evidently, the peak wavelength gets redshifted on increasing particle size R in the dimer configuration. This provides a mechanism for spectral fine tuning through control of nanoparticle size.

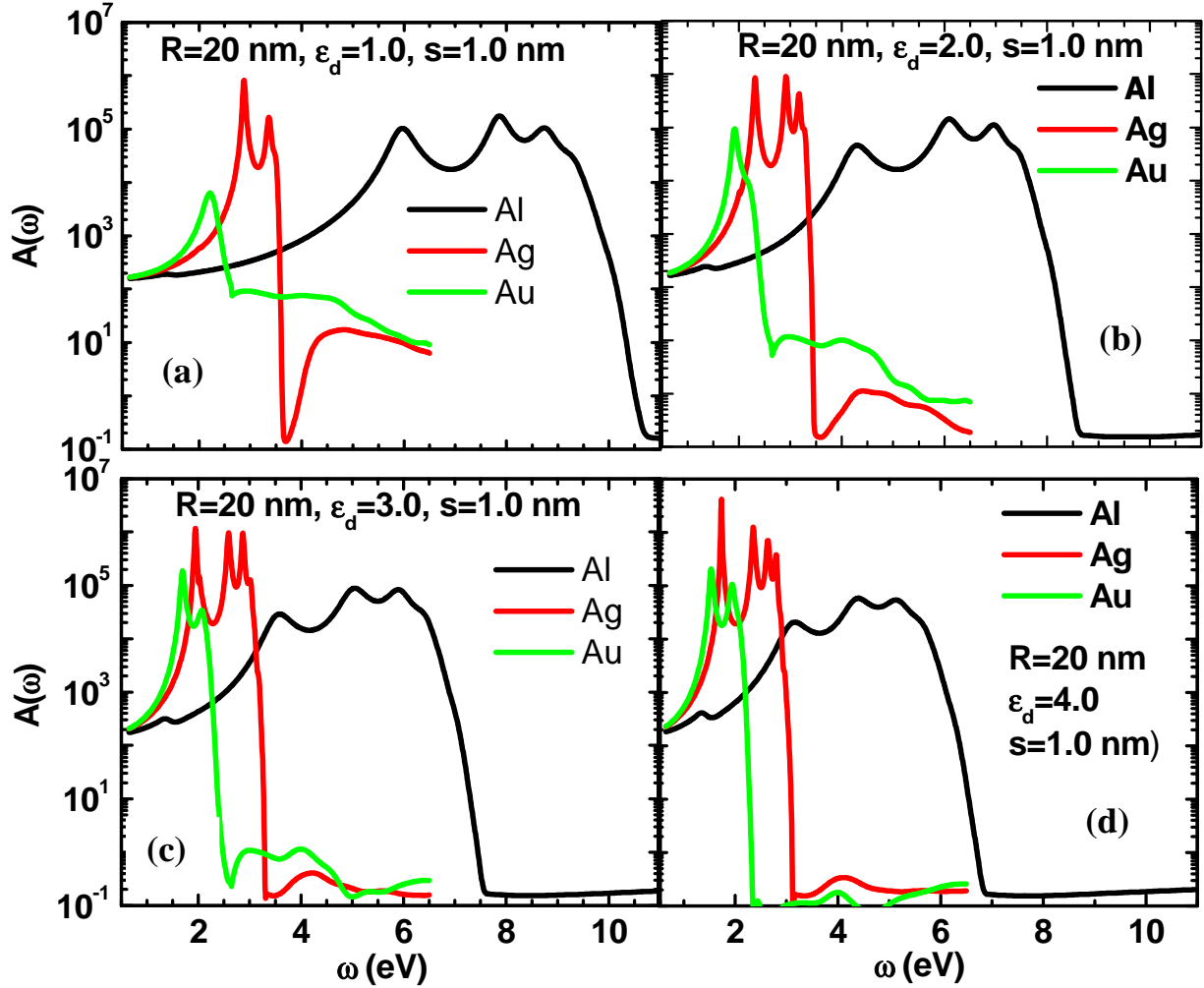

**Fig. 5: (a).** The variation of A(ω) with photon energy ℏω for dimer configurations of Al, Ag and Au having R=20 nm and varying embedding mediums, $\varepsilon_d$=1.0 **(a)**, 2.0 **(b)** 3.0 **(c)** and 4.0 **(d)**. Interparticle separation, S in this case is taken 1.0 nm. A clear redshift can be observed in all the cases. Multiple resonance peaks appears in the spectra of all three metals.

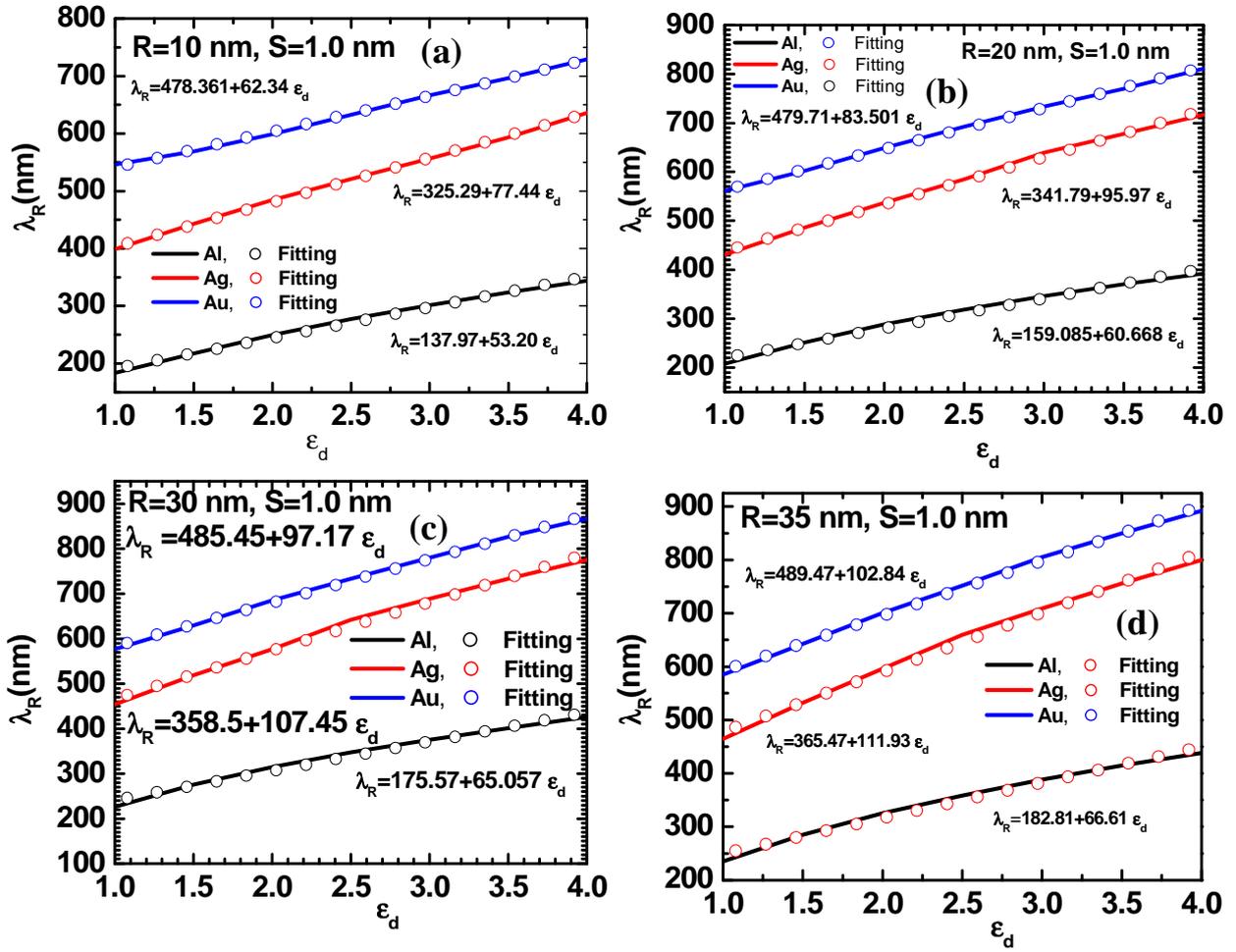

**Fig.6:** Calculated dependence of resonant frequencies $\lambda_R$ (for first mode) on the dielectric permittivity of the embedding medium for Al, Ag, and Au with nanoparticle radius=10-nm (a), R=20 nm (b), R=30 nm (c) and R=35 nm (d). For all calculations, S=1.0 nm is considered. It can be seen that the resonant position gets redshifted with increasing dielectric constant of embedding medium. All calculated data is also fitted and shown as open circles. Evidently, Al shows enhanced response in U.V. region and Ag/Au in Visible spectral region.

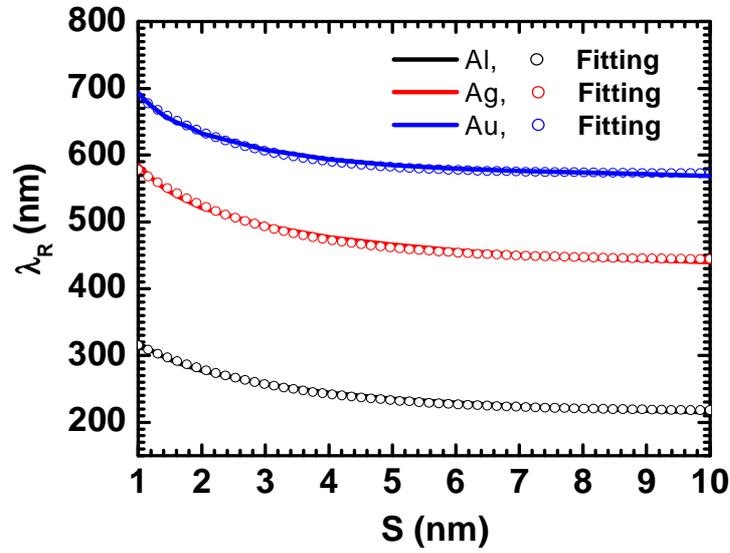

**Fig. 7:** Calculated dependence of resonant wavelength, $\lambda_R$ on interparticle separation, S for Al, Ag, and Au plasmonic materials. For these calculations, R=20 nm, and $\varepsilon_d$=2.5 is considered. Evidently, resonant wavelength gets blueshifted with increasing interparticle separation S. The variation of $\lambda_R$ with S is fitted as an exponentially decaying function characterized by characteristic decay separation $S_0$. The characteristic separation is maximum for Al and minimum for Au. It can be seen that, the resonant position tends toward a fixed value beyond certain interparticle separation.

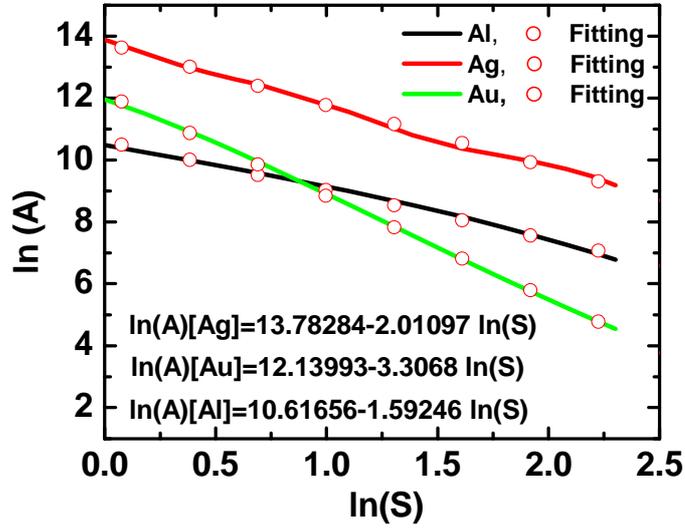

**Fig. 8:** Calculated dependence of field amplification factor (A) on interparticle separation (S) for Al, Ag, and Au. For these calculations, R=20 nm, and $\varepsilon_d=2.5$ is considered. Evidently, the amplification factor reduces drastically on increasing interparticle separation S. The rate of field reduction is maximum for Au and minimum for Al. Also shown is the fitted data and expression for all cases. The characteristic distance is maximum for Al and minimum for Au.